\documentclass[sigconf]{acmart}

\AtBeginDocument{%
  \providecommand\BibTeX{{%
    \normalfont B\kern-0.5em{\scshape i\kern-0.25em b}\kern-0.8em\TeX}}}

\copyrightyear{2023}
\acmYear{2023}
\setcopyright{acmlicensed}
\acmConference[CIKM '23]{Proceedings of the 32nd ACM International Conference on Information and Knowledge Management}{October 21--25, 2023}{Birmingham, United Kingdom} \acmBooktitle{Proceedings of the 32nd ACM International Conference on Information and Knowledge Management (CIKM '23), October 21--25, 2023, Birmingham, United Kingdom}
\acmPrice{15.00}
\acmDOI{10.1145/3583780.3615243} 
\acmISBN{979-8-4007-0124-5/23/10}

\usepackage{multirow}
\usepackage{graphicx}
\usepackage{balance}

\renewcommand{\footnoterule}{
  \kern -8pt                        
  \hrule 
  \kern 2pt}                   

\begin{document}

\title{MvFS: Multi-view Feature Selection for Recommender System}

\captionsetup[table]{skip=2pt}
\captionsetup[figure]{skip=2pt}
\newcommand{\proposed}{MvFS\xspace}
\newlength{\textfloatsepsave} \setlength{\textfloatsepsave}{\textfloatsep} \setlength{\textfloatsep}{0.5pt}

\author{Youngjune Lee}
\affiliation{%
  \institution{NAVER Corporation, South Korea}
  \country{}
}
\email{youngjune.lee93@navercorp.com}

\author{Yeongjong Jeong}
\affiliation{%
  \institution{NAVER Corporation, South Korea}
  \country{}
}
\email{yeongjong.jeong@navercorp.com}

\author{Keunchan Park}
\affiliation{%
  \institution{NAVER Corporation, South Korea}
  \country{}
}
\email{keunchan.park@navercorp.com}

\author{SeongKu Kang}
\affiliation{%
  \institution{University of Illinois at Urbana-Champaign, USA}
  \country{}
}
\authornote{SeongKu Kang is the corresponding author.}
\email{seongku@illinois.edu}

\renewcommand{\shortauthors}{Youngjune Lee, Yeongjong Jeong, Keunchan Park, \& SeongKu Kang}
\begin{abstract}
Feature selection, which is a technique to select key features in recommender systems, has received increasing research attention. 
Recently, Adaptive Feature Selection (AdaFS) has shown remarkable performance by adaptively selecting features for each data instance, considering that the importance of a given feature field can vary significantly across data.
However, this method still has limitations in that its selection process could be easily biased to major features that frequently occur.
To address these problems, we propose Multi-view Feature Selection (MvFS), which selects informative features for each instance more effectively. 
Most importantly, MvFS employs a multi-view network consisting of multiple sub-networks, each of which learns to measure the feature importance of a part of data with different feature patterns. 
By doing so, MvFS mitigates the bias problem towards dominant patterns and promotes a more balanced feature selection process. Moreover, MvFS adopts an effective importance score modeling strategy which is applied independently to each field without incurring dependency among features. Experimental results on real-world datasets demonstrate the effectiveness of MvFS compared to state-of-the-art baselines.
\end{abstract}


\maketitle

\section{Introduction}
In real-world web platforms, recommender systems (RS) encounter a multitude of features collected from users (e.g., age, gender), items (e.g., brand, price), and their interactions (e.g., dwell time, location), and the effective utilization of these features plays a critical role in the quality of recommendations. 
To capture the intricate relationships of the features, RS models have employed sophisticated architectures with powerful capacities \cite{wang2017dcn, guo2017deepfm, DERRD, HetComp, concf, TD}. 
However, it has been noted that some of these features lack relevance or exhibit redundancy in the context of user-item interactions, and thereby blindly feeding all features into the model often leads to suboptimal accuracy and slower model optimization \cite{lin2022adafs, wang2022autofield, lyu2023optimizing, lyu2022optembed}.
To address this problem, feature selection, which is a technique to select a subset of the most informative features, has recently received increasing attention \cite{liu2021mining,liu2020autofis, wang2022autofield, lyu2023optimizing, lin2022adafs,lyu2022optinter}.

In the past decades, a variety of methods has been studied for feature selection, ranging from hand-crafted and statistical methods \cite{wold1987principal, yang1997comparative, hoerl1970ridge, hall1999correlation} to traditional machine learning methods \cite{natekin2013gradient,tibshirani1996regression}.
However, they often show limited efficacy as the feature selection is conducted independently from the subsequent RS model, disregarding the model's prediction behaviors \cite{lyu2023optimizing}.
Recent methods \cite{lin2022adafs, wang2022autofield, lyu2023optimizing} have employed the AutoML approach \cite{liu2018darts,luo2018NAS} to automatically identify and select the most predictive features during the model optimization, and have shown effectiveness for RS models.

A state-of-the-art method, Adaptive Feature Selection (AdaFS) \cite{lin2022adafs}, introduces a new strategy that adaptively selects features tailored to each user-item interaction pair, considering that the importance of a given feature field can vary significantly across pairs. To handle such dynamic nature of each pair, AdaFS employs a controller network that computes the importance of each feature field for each pair. Then, it generates weighted feature embeddings by multiplying the importance. With the adaptive selection, AdaFS significantly improves performance over the previous methods that aim to select a \textit{globally fixed} subset of feature fields \cite{wang2022autofield, tibshirani1996regression, wold1987principal, natekin2013gradient}.

Still, there is much room for improvement in AdaFS.
First, it employs a single controller network to select features for all user-item pairs.
This makes the selection process could be easily biased in favor of a few large groups of pairs having features that occur frequently, consequently resulting in limited improvement for data with minor features having a relatively low frequency.
We argue that the efficacy of the adaptive selection can be further improved by explicitly preventing the bias problem.
Second, AdaFS applies a reweighting step to ensure that the sum of the computed importance remains constant before being reflected in the subsequent RS model. 
However, this reweighting introduces fluctuations in the overall scale of importance scores according to the number of selected fields, e.g., with $k$ selected fields, the average importance scale is proportional to $1/k$, which creates unnecessary dependencies among the features resulting in suboptimal performance.

In this work, we propose the multi-view feature selection (MvFS), which aims to address the aforementioned shortcomings.
To this end, MvFS first employs multiple sub-networks within the controller. 
Each sub-network in MvFS is designed to specialize in data with different feature patterns, preventing it from being biased to a few dominant patterns and promoting a more balanced feature selection process. 
Moreover, MvFS adopts a new importance score modeling strategy that is applied independently to each field without the reweighting step. We validate the superiority of MvFS with extensive experiments on real-world datasets, and provide a detailed analysis showing the effectiveness of each proposed component.

\section{Methodology}
We first provide the overview of RS model with feature selection (Sec.2.1).
Then, we present our multi-view feature selection strategy (Sec.2.2) and optimization procedure (Sec.2.3). 

\begin{figure}
\centering
  \includegraphics[width=56mm, height = 46mm]{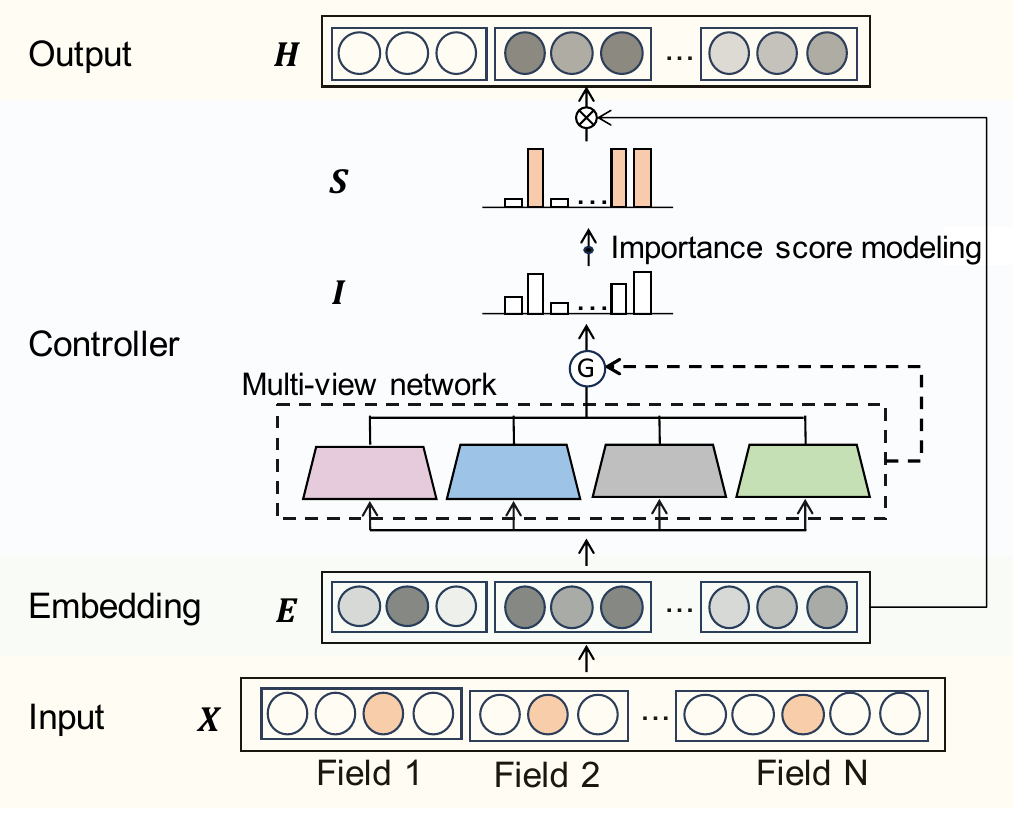}
  \caption{Overview of MvFS.
  }
  \label{fig:overview}
\end{figure}

\subsection{RS model with Feature Selection}
In this work, we focus on the adaptive feature selection that adaptively selects informative feature fields for each user-item pair \cite{lin2022adafs}.

\subsubsection{\textbf{Input construction.}}
The raw data consists of several feature fields (e.g., gender, price). 
The most common way to construct inputs for the RS model is embedding lookup \cite{lyu2022optembed, song2019autoint, guo2017deepfm, SSCDR, wei2021autoias}.
Formally, given $N$ feature fields, we denote each data instance as $\boldsymbol{X}$ = [$\boldsymbol{x_1}, ..., \boldsymbol{x_N}$], where $\boldsymbol{x_n}$ is a one-hot vector that encodes the value of $n^{th}$ feature field. 
Let $\boldsymbol{P_n}$ denote a learnable projection matrix of $n^{th}$ field.
After projecting each feature field (i.e., the embedding lookup), the data instance $\boldsymbol{X}$ is converted to the dense vector $\boldsymbol{E}$ as: 
 \begin{equation}
     \boldsymbol{E} = [\boldsymbol{e_1}, \boldsymbol{e_2}, ..., \boldsymbol{e_N}],
 \end{equation}
where $\boldsymbol{e_n}$ = $\boldsymbol{P_n} \boldsymbol{x_n}$ denotes the embedding of each field. 

\subsubsection{\textbf{RS model with adaptive feature selection.}}
To selectively exploit informative features for each data, a controller network is employed to compute the importance score of each feature field.
Then, a weighted feature vector is generated by multiplying the computed importance score with the corresponding feature embedding.
The weighted feature vector $\boldsymbol{H}$ is obtained by
\begin{equation}
 \boldsymbol{H} = [s_1\boldsymbol{e_1}, s_2\boldsymbol{e_2}, ..., s_N\boldsymbol{e_N}],
\end{equation}
where $s_n$ is the importance score of $n^{th}$ field computed by the controller.
$s_n$ can be either a real value (soft selection) or a binary value (hard selection), depending on the design choice \cite{lin2022adafs}.
Lastly, the weighted feature vector is fed into the subsequent RS model:
\begin{equation}
     \hat{y} = RS(\boldsymbol{H}),
\end{equation}
where $\hat{y}$ is the model prediction.
$RS(\cdot)$ can be any recommendation model, and it is treated as a black box.
It is worth noting that feature selection only affects the model input, allowing it to be flexibly applied to various architectures.

\subsection{Multi-view Feature Selection}
We present Muiti-view Feature Selection (MvFS) with a new controller designed to select informative features while avoiding bias toward a few dominant feature patterns (Figure \ref{fig:overview}).
Our controller consists of two components: (1) multi-view network which computes feature importance by incorporating various views from multiple sub-networks, and (2) importance score modeling which decides the final importance score for the feature selection.

\subsubsection{\textbf{Multi-view network}}
Multi-view network computes the importance of each feature field by taking the feature vector $\boldsymbol{E}$ as input.
The prior method \cite{lin2022adafs} employs a single network for computing feature importance, and this makes the controller could be easily biased towards a few major features that occur frequently.
To tackle this problem, we adopt the idea of the mixture-of-experts \cite{shazeer2017moe,jacobs1991adaptive}.
It exploits a divide-and-conquer strategy composed of multiple distinct sub-networks, each of which learns to handle a part of input space.
By leveraging multiple sub-networks focusing on data with different feature patterns, we aim to prevent 
to the controller from being biased and enable a more balanced feature selection.

Concretely, the multi-view network consists of (1) $K$ sub-networks ($SN_1, SN_2, ..., SN_K$) quantifying the importance of feature fields, and (2) a gating module $g$ that regulates the influence of each sub-network based on the feature patterns of each data. 
We first calculate the feature importance by $SN_k(\boldsymbol{E}) \in \mathbb{R}^N$, which yields an importance vector where each element represents the importance of each field.
Then, the influences of the sub-networks $\boldsymbol{r} \in \mathbb{R}^K$ are computed using $g$ as follows:
\begin{equation}
\begin{aligned}
     \boldsymbol{r} = \sigma(\boldsymbol{W_g C} + \boldsymbol{b_g}), \quad  \boldsymbol{C} = [SN_1(\boldsymbol{E}), ... , SN_K(\boldsymbol{E})],
\end{aligned}
 \end{equation}
where $\sigma$ is the sigmoid function, $\boldsymbol{W_g} \in \mathbb{R}^{K \times KN}$ and $\boldsymbol{b_g} \in \mathbb{R}^{K}$ are learnable weight matrix and bias vector of the gating module.
We use the outputs of the sub-networks for gating so that we can use the summarized information on the input feature vector\footnote{We empirically obtained better results compared to the case of using the feature vector itself as input of the gating module, i.e.,  $C=\boldsymbol{E}$.}.
The data sharing similar feature patterns would naturally have similar outputs from the sub-networks, resulting in similar gating results.

The final feature importance vector $\boldsymbol{I} \in \mathbb{R}^{N}$ is computed by aggregating the output of the sub-networks based on the gating results as follows:
\begin{equation}
\begin{aligned}
    \boldsymbol{I} =  \sum_{k} r_k \cdot SN_k(\boldsymbol{E}),
      \end{aligned}
\label{importance}
 \end{equation}
where $r_k$ denotes the $k^{th}$ value in $\boldsymbol{r}$.

\begin{table*}[ht]
\small
\caption{Performance comparison of feature selection methods. * indicates $p$ $\le$ 0.05 for the two-sided t-test over the best baseline.}
\centering
\begin{tabular}{c|c|cc|cc|cc|cc|cc}

\hline
\multirow{2}{*}{Dataset}   & \multirow{2}{*}{RS Model}  & \multicolumn{2}{c|}{No Selection}   & \multicolumn{2}{c|}{AutoField} & \multicolumn{2}{c|}{AdaFS} & \multicolumn{2}{c|}{OptFS} & \multicolumn{2}{c}{MvFS}\\ \cline{3-12}
  & & AUC $\uparrow$ & Logloss $\downarrow$ & AUC $\uparrow$ & Logloss $\downarrow$  & AUC $\uparrow$ & Logloss $\downarrow$  & AUC $\uparrow$ & Logloss $\downarrow$ & AUC $\uparrow$ & Logloss $\downarrow$   \\ \hline\hline
\multirow{4}{*}{Avazu}  & MLP   &  0.7823  & 0.3794  &  0.7823  & 0.3791  &  0.7832  & 0.3793  &  0.7814  & 0.3804 & \textbf{0.7866*} & \textbf{0.3765*}\\ 
                     & DeepFM   & 0.7836     & 0.3781   &  0.7830  & 0.3786 &  0.7835  & 0.3783 &  0.7850  & 0.3771 & \textbf{0.7871*} & \textbf{0.3761*}\\ 
                     & DCN  & 0.7839    & 0.3779 &  0.7841  & 0.3778 &  0.7854  & 0.3767 &  0.7859  & 0.3764& \textbf{0.7884*}& \textbf{0.3748*} \\ 
                    & IPNN       & 0.7864    & 0.3765  &  0.7862  & 0.3766 &  0.7837  & 0.3790 &  0.7861  & 0.3767 & \textbf{0.7898*} & \textbf{0.3745*}  \\ \hline
\multirow{4}{*}{Criteo}  & MLP   &  0.8040  & 0.4481  &  0.8038  & 0.4494  &  0.8083  & 0.4435  &  0.8027  & 0.4490 & \textbf{0.8107*} & \textbf{0.4412*}\\ 
                     & DeepFM   &  0.8094  & 0.4424  &  0.8095  & 0.4425  &  0.8086  & 0.4439  &  0.8043  & 0.4469 & \textbf{0.8118*} & \textbf{0.4401*}\\ 
                     & DCN  &  0.8059  & 0.4455  &  0.8060  & 0.4454  &  0.8084  & 0.4435  &  0.8065  & 0.4451 & \textbf{0.8116*} & \textbf{0.4403*} \\ 
                    & IPNN      &  0.8093  & 0.4428  &  0.8093  & 0.4426  &  0.8083  & 0.4440  &  0.8082  & 0.4433 & \textbf{0.8111*} & \textbf{0.4413*} \\                     \toprule
\end{tabular}
\label{table:performance}
\vspace{-0.4cm}
\end{table*}

\subsubsection{\textbf{Importance score modeling}} 
We utilize the computed importance of each field to model the final importance scores. 
To strike a balance between exploration and exploitation, we employ a gradual transition from soft selection to hard selection during the training process. 
In the early stages, the RS model explores various feature combinations through soft selection. 
As training progresses, we gradually prioritize informative features while disregarding unimportant and redundant ones through hard selection. 
The final importance score for the $n^{th}$ feature field is defined as follows:
 \begin{equation}
 \begin{aligned}
    s_n =  0.5 * (1 + {\tanh({\tau \cdot (I_n - l )})}),
  \end{aligned}
 \end{equation}
\noindent
where $l$ is the threshold for selection. 
We model the transition using an approximation of the unit step function with the $\tanh$ function. 
We set $\tau = max(5, 1+0.001t)$ to control the smoothness of the $\tanh$, where $t$ denotes the training step.
This choice gradually makes $s_n$ a binary value, enabling a smooth transition to hard selection during the training.
It is worth noting that our score modeling is applied independently to each field, unlike the prior method \cite{lin2022adafs} that uses the reweighting step across the fields which incurs unnecessary dependencies among the selected features.

Lastly, we apply feature selection by multiplying the importance scores $\boldsymbol{s}$ with the feature embedding $\boldsymbol{E}$, i.e., $\boldsymbol{H} = [s_1\boldsymbol{e_1}, s_2\boldsymbol{e_2}, ..., s_N\boldsymbol{e_N}]$.
At the test time, we apply the hard selection with the step function.

\subsection{Optimization}
We train the controller and the RS model to predict interactions of user-item pairs.
We denote the weighted feature vector for the $m^{th}$ instance as $\boldsymbol{H}^m$, and the corresponding ground truth label as $y^m$.
The loss function is defined as follows:
\begin{equation}
 \begin{aligned}
    \min_{\boldsymbol{\theta_{RS}}, \boldsymbol{\theta_{C}}} \frac{1}{M}\sum_{m=1}^{M} \mathcal{L}_{BCE}(RS(\boldsymbol{H}^m),y^m),
 \end{aligned}   
\end{equation}
\noindent
where $\boldsymbol{\theta_{RS}}$ denotes the parameters of the RS model, including the embedding components and the subsequent layers.
$\boldsymbol{\theta_{C}}$ denotes the parameters of the controller.
To ensure reliable feature embeddings for feature selection, we initially warm up the parameters of the RS model for a few epochs without using the controller \cite{lin2022adafs}.

\section{Experiments}

\subsection{Experiment Setup}

\noindent
\textbf{Datasets.} We use two public real-world benchmark datasets: Avazu\footnote{https://www.kaggle.com/c/avazu-ctr-prediction} and Criteo\footnote{https://ailab.criteo.com/ressources/}. 
We randomly split each dataset 8:1:1 for training, validation, and testing.

\noindent
\textbf{Metrics.}
We use AUC score and Logloss, which are widely used for CTR prediction task, as evaluation metrics. 
Note that a \textbf{0.001-level} increase in AUC indicates a significant improvement \cite{guo2017deepfm,cheng2016wide}.

\noindent
\textbf{Baselines \& RS models.} 
We compare the MvFS with the following feature selection methods: (1) \textbf{AutoField} \cite{wang2022autofield} selects globally fixed feature fields using neural architecture search techniques \cite{liu2018darts}.
(2) \textbf{AdaFS} \cite{lin2022adafs} adaptively selects feature fields for each data instance via the controller (Sec.2.1). 
(3) \textbf{OptFS} \cite{lyu2023optimizing} selects informative global features considering feature interactions. 
We evaluate the efficacy of the feature selection on the following RS models: MLP \cite{zhang2016deep}, DeepFM \cite{guo2017deepfm}, DCN \cite{wang2017dcn}, and IPNN \cite{qu2016ipnn}.

\noindent
\textbf{Implementation details.}
Our implementation\footnote{\url{https://github.com/dudwns511/MvFS_CIKM23}} is based on the public PyTorch library for RS\footnote{\url{https://github.com/rixwew/pytorch-fm}}.
We utilize the official implementation for AdaFS\footnote{\url{https://github.com/Applied-Machine-Learning-Lab/AdaFS}} and OptFS\footnote{\url{https://github.com/fuyuanlyu/OptFS}}.
We set the embedding dimension as 16 and the batch size as 4096. 
For all RS models, we set the MLP layer as two fully-connected layers of size [16, 8]. 
For all compared methods, we tune the learning rate and L2 regularizer from \{1e-3, ... , 1e-6\}. 
For MvFS, the number of sub-network $K$ is selected from \{3, 4, 5\}, and the selection threshold $l$ is selected from \{0.1, 0.2, 0.3, 0.4\}.
For the sub-network $SN_k$, we employ a simple linear layer with Softmax activation.
We set the warm-up epoch to~3.

\subsection{\textbf{Performance Comparison}}
Table \ref{table:performance} presents the performance of various feature selection methods on four different RS models.
First, AutoField and OptFS show a limited improvement over the base RS model (i.e., No Selection), and they often show degraded performance.
This result shows the limitation of globally fixed feature selection that cannot consider varying feature importance for each data instance.
Second, AdaFS, which adaptively selects the features for each data instance, shows performance gains to some extent. 
However, with a single network-based controller, its selection process could be easily biased to a few dominant features. 
Lastly, MvFS shows significant improvement over other baselines with the multi-view network which computes feature importance by incorporating various views from multiple sub-networks, which can effectively prevent the bias problem.

\begin{table}[ht]

\small
\caption{Transferability analysis on Avazu. 
* indicates $p$ $\le$ 0.05 for the two-sided t-test over the No Selection.}
\centering

\begin{tabular}{c|ll|ll}
\hline
\multirow{2}{*}{RS Model}  & \multicolumn{2}{c|}{No Selection} & \multicolumn{2}{c}{Transferred MvFS}  \\ \cline{2-5}
 &  AUC $\uparrow$ & Logloss $\downarrow$ & AUC $\uparrow$ & Logloss $\downarrow$  \\ \hline\hline
 DeepFM      & 0.7836  & 0.3781 & \textbf{0.7864*} &\textbf{0.3765*}  \\
 DCN      & 0.7839  & 0.3779 & \textbf{0.7875*}  &\textbf{0.3759*}  \\
 IPNN      & 0.7864  & 0.3765 & \textbf{0.7869}  &\textbf{0.3761}  \\\toprule

\end{tabular}
\label{table:transfer}
\end{table}

\subsection{\textbf{Study of MvFS}} 

\noindent
\textbf{Transferability analysis.}
We conduct an analysis to validate the transferability of MvFS. 
Following \cite{lin2022adafs}, we first train MvFS with the MLP model and freeze the controller parameters.
Then, we utilize the controller for feature selection of other RS models (i.e., DeepFM, DCN, and IPNN). 
The results are presented in Table \ref{table:transfer}. 
All models achieve higher performance with the transferred controller of MvFS, which support that MvFS can consistently select the most informative features for different RS models.
It is worth noting that the best performance of each RS model is achieved when MvFS is jointly trained with each model (Table \ref{table:performance}).
This shows that the RS models have different prediction behaviors, and thereby feature selection needs to be tailored for each model, as discussed in \cite{lyu2023optimizing}.

\begin{figure}[h]
\centering
  \includegraphics[width=74mm, height = 29mm]{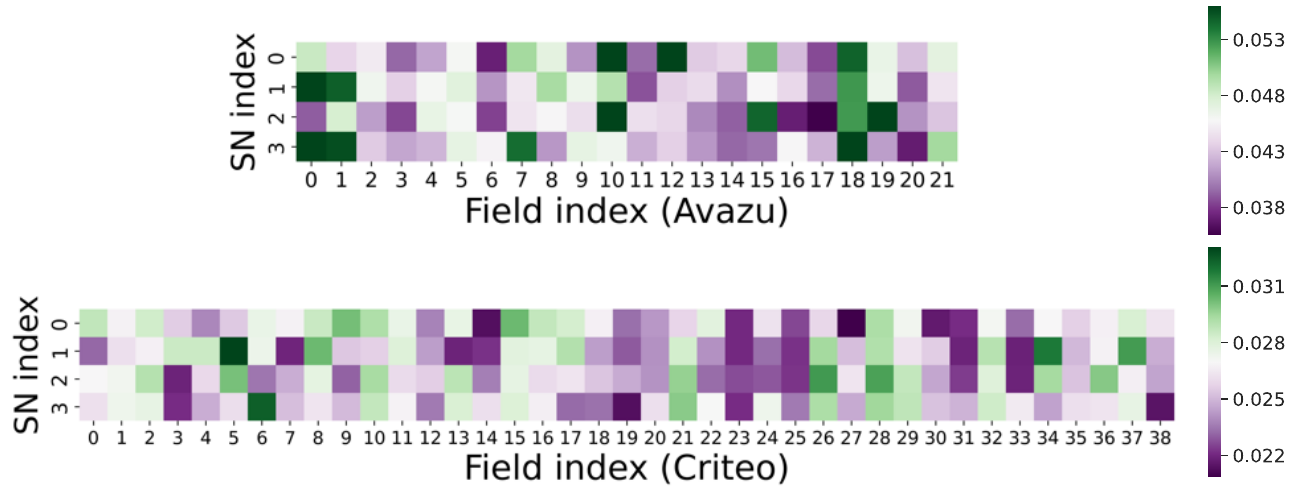}
  \caption{The average feature importance computed by each sub-network. Results with MLP.}
  \label{fig:importance}
 \vspace{-0.4cm}
\end{figure}

\begin{figure}[h]
\centering
  \includegraphics[width=68mm, height=23mm]{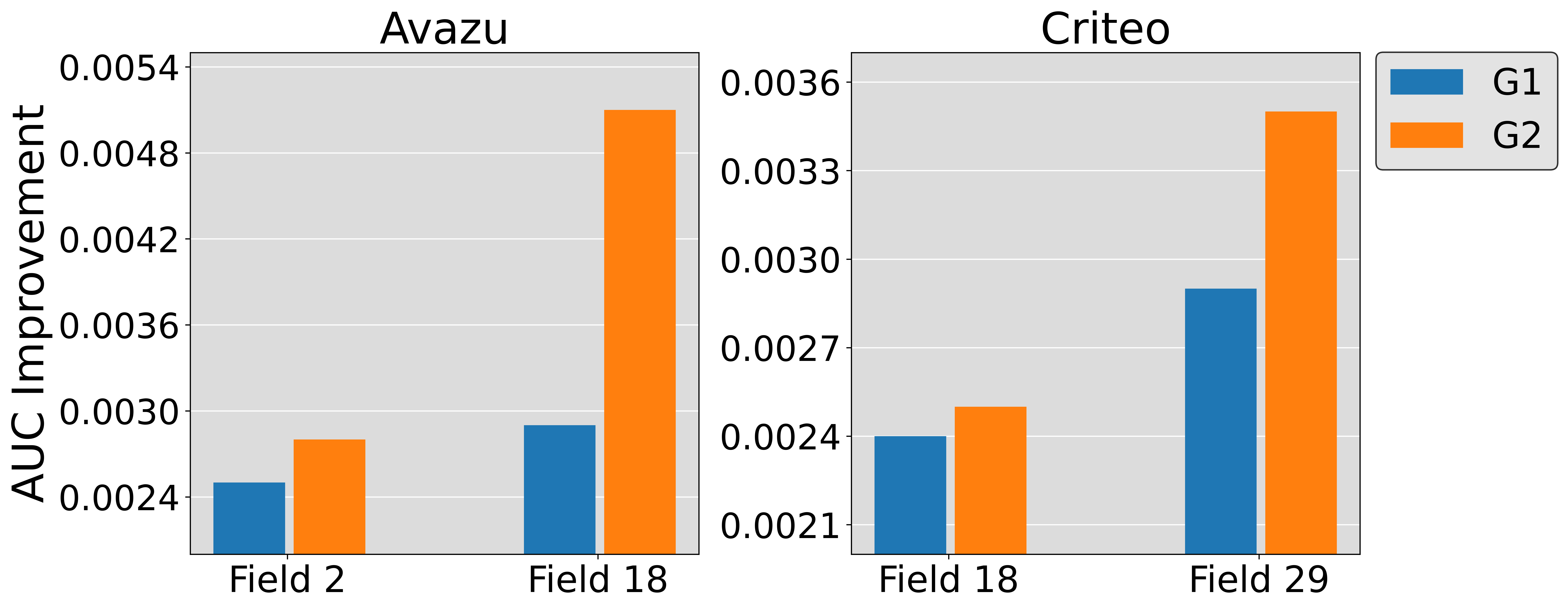} 

  \caption{AUC Improvement over AdaFS.
  (Left) Field 2 and Field 18 of Avazu.
  (Right) Field 18 and Field 29 of Criteo. 
  For each dataset, we select two fields that AdaFS shows the most limited performance in the minor feature group (G2).}
  \label{fig:case}
  \vspace{-0.1cm}
\end{figure}

\noindent
\textbf{Effectiveness of controller.}
We provide empirical analysis to support the effectiveness of our multi-view controller from two perspectives: \textbf{(a)} Whether each sub-network meaningfully captures different feature patterns. \textbf{(b)} Whether MvFS selects features in a more balanced way than a single network (i.e., AdaFS). 

For \textbf{(a)}, we analyze the differences in the output of sub-networks in Figure \ref{fig:importance}. 
In specific, we compute the sum of each sub-network output (i.e., $r_k \cdot SN_k(\boldsymbol{E})$ of Eq.\ref{importance}) on all test data instances and divide by the total number of instances. 
We observe that each sub-network assigns higher importance to different feature fields.
For instance, on Avazu, the sub-network with index 0 tends to consider fields 10, 12, and 18 as important, while the sub-network with index 1 also considers fields 0, 1, and 18 as important.
This result shows that each sub-network indeed focuses on different parts of data having distinct feature patterns.

For \textbf{(b)}, we analyze the improvement by MvFS on data with minor features having a relatively low frequency. 
To this end, we split the data into two groups according to the feature frequency: (1) G1 with the top 95\% frequent features and (2) G2 with the remaining ones.
Then, we present the absolute improvement over AdaFS for each group in Figure \ref{fig:case}. 
We observe that MvFS achieves higher performance than AdaFS on both G1 and G2, with a particularly significant improvement in G2. 
This result shows that our multi-view controller indeed better prevents it from being biased to the major features and enables a more balanced feature selection.
Further, this result supports the superior performance of MvFS~in~Table~\ref{table:performance}.

\begin{figure}[h]
\centering
  \includegraphics[width=55mm, height=21.5mm]{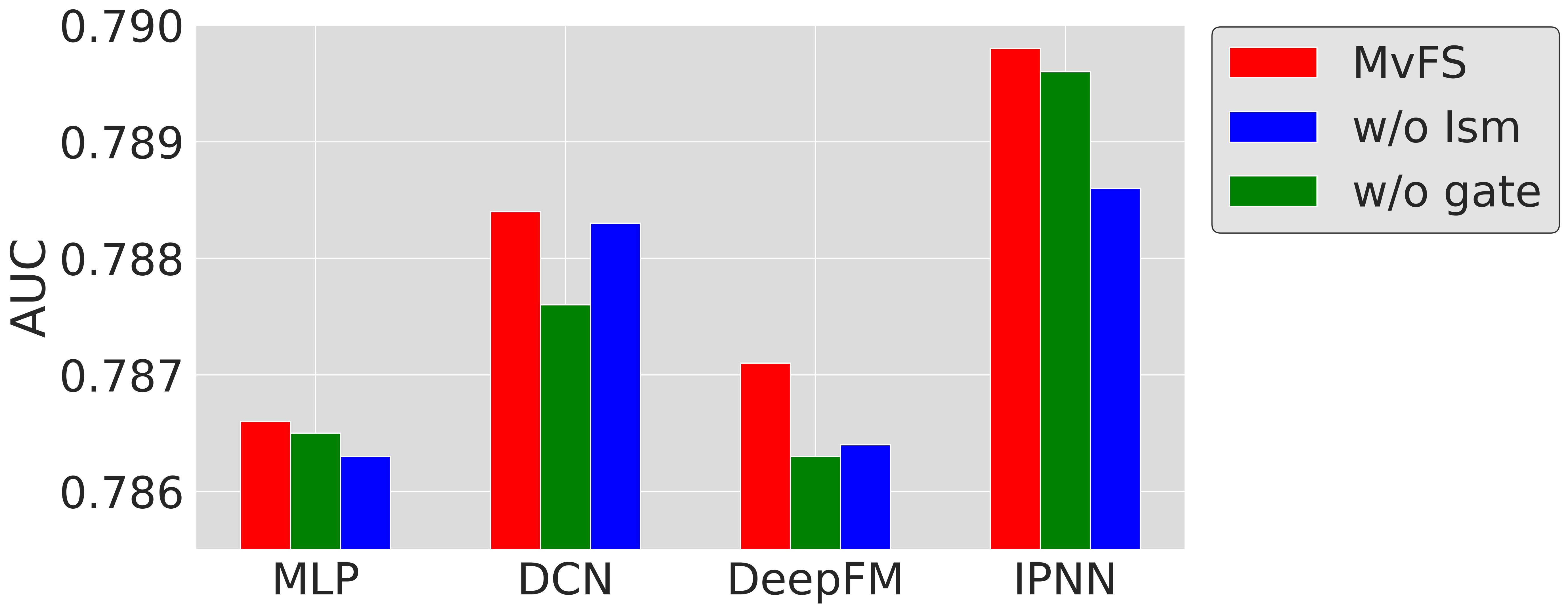}
  \caption{Ablation study on Avazu.}
  \label{fig:ablation}
\end{figure}
  
\noindent
\textbf{Ablation study.}
We provide an ablation study of two key components of MvFS in Figure \ref{fig:ablation}. 
We compare the following ablations:
(1) \textbf{w/o Ism} ablates the importance score modeling (Sec.2.2.2). 
It uses $I$ (Eq.\ref{importance}) for the feature selection.
(2) \textbf{w/o gate} ablates the gating module which controls the influence of each sub-network.
It uses a uniform distribution for $r_k$.
We observe that each component brings significant performance improvements across the different RS models.
These results support the superiority of our strategy that controls the scale of importance scores without incurring dependency among the selected features (\textbf{w/o Ism}), and our strategy that controls the influences of sub-networks based on the feature pattern of each data (\textbf{w/o gate}).

\begin{figure}[h]
\centering
  \includegraphics[width=37mm, height=22.5mm]{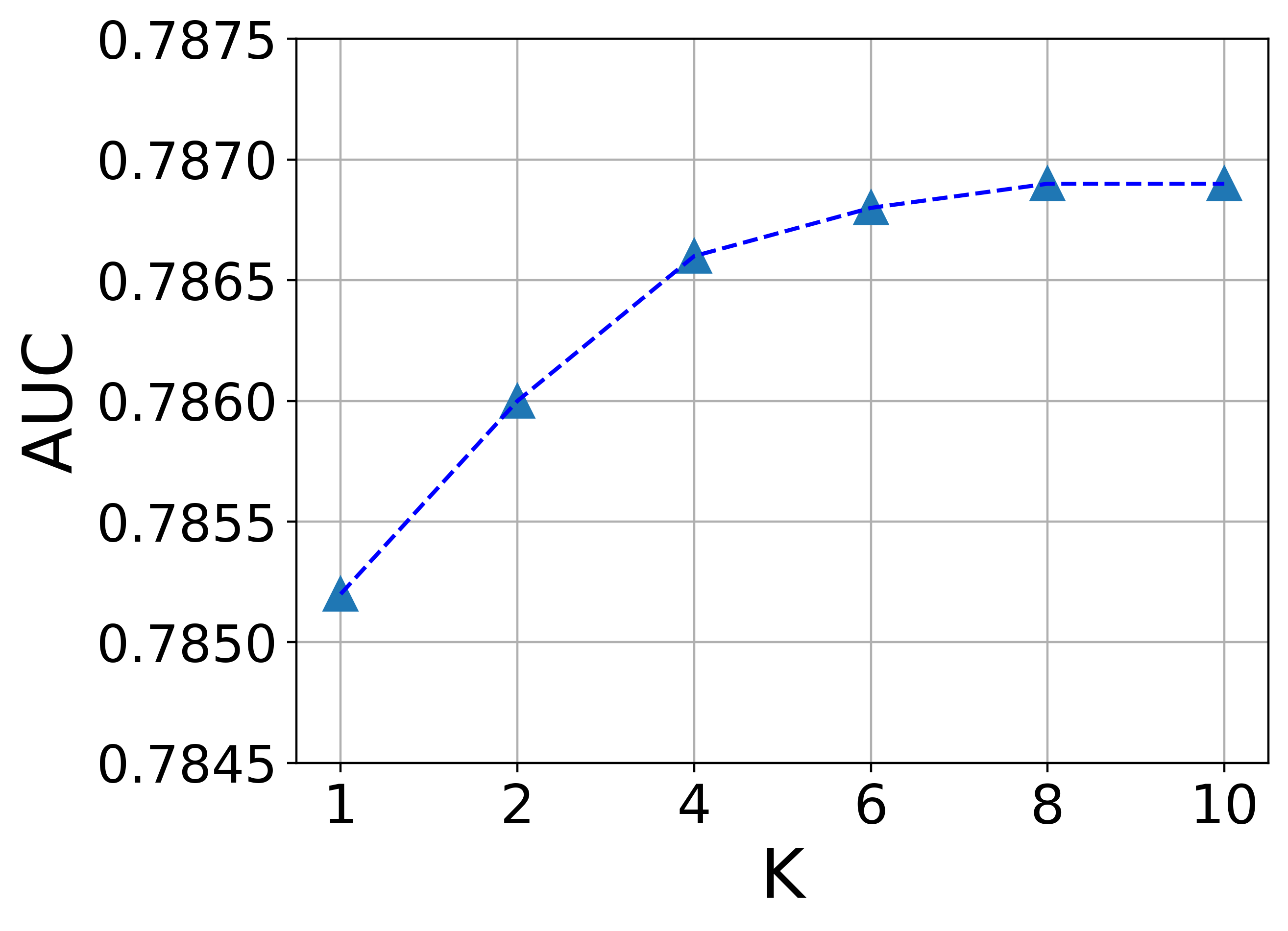}
  \includegraphics[width=37mm, height=22.5mm]{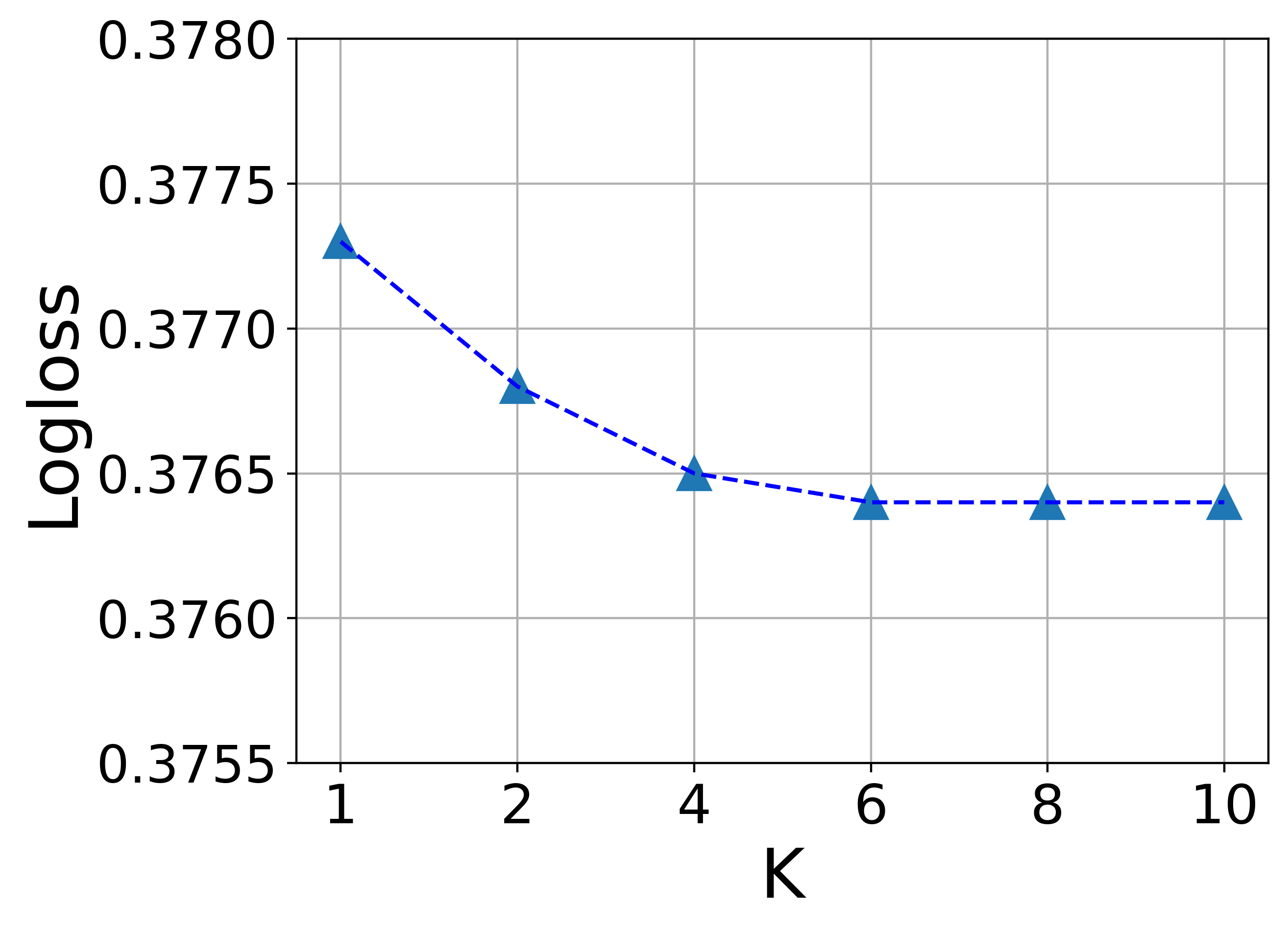}
  \caption{Effects of the number of sub-network.
  (Left) AUC,
  (Right) Logloss results of MLP on Avazu.}
  \label{fig:param}
\end{figure}

\noindent
\textbf{Hyperparameter Study.}
In Figure \ref{fig:param}, we report the performance of MvFS with varying $K$, the number of sub-networks in the controller.

Compared to the case of a single sub-network ($K=1$), the multi-view network ($K\geq2$) brings significant performance improvement.
Also, the performance tends to converge as $K$ increases. The best performance is achieved near $6-8$.

\section{Conclusion}
We propose MvFS, a multi-view feature selection that selects informative features for each data instance effectively. 
MvFS uses adaptive feature selection with the multi-view network which computes feature importance by incorporating various views from multiple sub-networks, enabling a more balanced feature selection.
Moreover, MvFS adopts a new importance
score modeling strategy that computes importance scores without incurring unnecessary dependencies among the selected features. 
We conduct extensive experiments to validate the effectiveness of MvFS on real-world datasets. 
Also, we conduct an in-depth study to ascertain the effectiveness of our multi-view approach.

\bibliographystyle{ACM-Reference-Format}
\balance
\bibliography{acmart}

\end{document}